\documentclass[aps,onecolumn,superscriptaddress,floatfix]{revtex4}

\usepackage{bm}
\usepackage{amsmath}
\usepackage{latexsym}
\usepackage{amssymb,amsmath,graphicx}
\usepackage{amsfonts}
\usepackage[caption=false]{subfig}
\usepackage{epsfig,latexsym,graphicx}
\usepackage{amssymb}
\usepackage{graphicx}

\newcommand{\beq}{\begin{equation}}
\newcommand{\eeq}{\end{equation}}

\usepackage{epsfig}

\begin{document}
\title{ 
Universal City-size distributions through  rank ordering  
}
\author{Abhik Ghosh}
	\affiliation{Interdisciplinary Statistical Research Unit, Indian Statistical Institute, Kolkata 700108}
 \email{abhianik@gmail.com}
	\author{Banasri Basu}
	\affiliation{Physics and Applied Mathematics Unit, Indian Statistical Institute, Kolkata 700108, India}
	\email{sribbasu@gmail.com}

\date{\today}

\begin{abstract}
We consider a two-parameter discrete generalized beta (DGB)  distribution and propose its universal applications to study the size-distribution of 
the urban agglomerations across various countries in the world, 
where the urban agglomerations include the small and mid-sized cities along with the heavily populated cities.  
Our proposition is validated by an exhaustive study  with the 3 decades' census data for India and China and census data of USA for a time window of 8 years.  
Moreover, we have studied the city size distributions for many different countries,  
like Brazil, Italy, Sweden, Australia, Uganda etc., from all the continents around the world according to the availability of the data. 
The detailed analyses exhibit a unique global  pattern for the city size distributions,  
from low to high size, across the world with various geographic and economic conditions. 
Further analyses  based on the entropy of the distribution provide insights on the underlying randomness and spreads of the city sizes within a country. 
The DGB distribution, a typical  rank order (RO) distribution, through its two parameters,  not only fits the data on wider range of  city sizes better than the well-known  power law for all the countries considered,
 it also helps us to characterize, discriminate and study their evolution over time.  
\end{abstract}
\maketitle

\section{Introduction}
\label{SEC:intro}

Cities are the examples of complex systems. 
However, the functioning of this type of complex system  depends upon social, economic, and environmental factors. 
In spite of the diversity of the physical forms of the cities, there is a pattern and order in the expansion of cities. 
The size and shape of the cities are known to play a fundamental role in social and economic life \cite{pnas, batty}.  
The study of the size distribution of urban areas in various parts of the world helps us to understand 
their dynamics  and also plays a key role to manage their growth and environmental impacts.

City size distribution for various countries has been extensively studied for several decades.  
In general it is believed  that the city sizes obey a remarkably simple law, known as Zipf’s law \cite {zipf} 
(alternatively known as Pareto distribution or simply power law) for cities with heavy population.  
Although a considerable amount of research   has been devoted to power law behaviors, in various contexts, however, 
when real data \cite{usa, brazil, kgbb, kgbb2013} is analyzed, 
in most of the cases the power law trend holds only for an intermediate range of values; 
there is a power law breakdown in the lower distribution tails \cite{stanley2000, newman2005}. 
This has motivated the studies on the size distribution of smaller cities in different countries  
\cite{12kgbb, 13kgbb, 14kgbb, 15kgbb, 16kgbb, 17kgbb, 18kgbb, 19kgbb,luck17I,luck17U}. 
Recently, it has been shown that  the data for the  Indian city size distribution exhibit 
a strong reverse Pareto in the lower tail, log-normal in the mid-range and a Pareto in the upper tail \cite{luck17I}.
A relatively simpler mixture of lognormal in the lower tail and Pareto in the upper tail has also been fitted 
for US city size distributions \cite{luck17U}. 
The existing literature on modeling the sizes of cities having small to extremely large populations
are mostly, up to the knowledge of the authors, if not all, confined to the Pareto law in the upper tail;
hence use of a more  complex mixture distribution is required 
(with another appropriate distribution in the lower or middle ranges).
Such mixture distributions are known to be difficult to fit statistically (estimation process is often complex)
and often difficult to interpret. There is also no evidence of one such mixture distribution providing significantly good fit 
to different types of city-size distributions in socio-economically different countries across the world. 

The goal of the present paper is to search for a global pattern of city size distributions along with their discrimination and evolution, so that one can compare all the cities in a universal framework and better understand the dynamics of urbanization in society. 


In this work, we focus on a particular two-parameter Rank-Ordered (RO) distribution,
namely the discrete generalized beta distribution (DGBD) \cite{Martinez-Mekler09}, for  cities 
with different  sizes in  a given country in any continent of the world, including the important economies like India,  China, USA, Brazil 
as well as the recent distribution of the sizes of  World Cities with heavy population.
This particular RO distribution is a discrete generalization of the continuous beta distribution, 
where a given property of a system of $N$ items is ordered according to its importance (rank)
so that the domain is the natural numbers 1 to $N$ instead of the interval $(0,1)$ of usual continuous beta distribution. 
Several  RO distributions including the DGBD and its further generalizations have already been studied to provide good fits
in the context of different count or rank-size data from arts and sciences \cite{Martinez-Mekler09,Ausloos/Cerqueti:2016,Alvarez-Martinez/etc:2011, Alvarez-Martinez/etc:2014, Oscar2017,
	Alvarez-Martinez/etc:2018, Lalit/etc:2018}. 
In the present paper, our main contribution is to propose and illustrate that the DGBD can be successfully 
applied to provide a simple, yet excellent, universal fit to the city size  data across different socio-economic countries. 
Our analyses show that the data for all countries under consideration follow, with a wonderful goodness-of-fit,  
the DGB distribution, a two parameter RO distribution,  incorporating the product of two power laws 
defined over the complete data set --  
one measured from left to right and another in the opposite direction.
For the vanishing value of one particular parameter,  
the DGB distribution simplifies to the usual Pareto law for the cities with heavy population (see Section \ref{SEC:ROD}). 
To understand the underlying process of the urban morphology, the Shannon entropy of the proposed DGB distribution has also been studied.
Based on the two parameters of the distribution and the corresponding entropy, 
one can indeed discriminate and study the evolution of city distributions in different countries across the world.

The rest of the paper is organized as follows. Sec.~II  deals with the mathematical formulation of the two-parameter DGBD. 
In the respective subsections A, B and C, we define our particular RO distribution and the corresponding Shannon entropy 
and  provide the methodology for the estimation of the parameters. 
In Sec.~II.D we detail the prediction and the goodness-of-fit test for examining the performance of our proposal.   
Sec.~III  describes the data used for our empirical analyses and also demonstrates the RO modeling of these data of different countries across the world, 
including India, China, USA, Brazil, Australia, 3 European countries and 3 African countries at different points of time.
This section also presents the analysis of the world-wide size distributions of cities and countries in the year 2018 (Sec.~IV.H).
Sec.~V narrates the discrimination and evolution of the city-size distributions across the countries
and the advantage of our proposal in this context. 
Finally, we conclude the paper with some discussions about the future works in Sec.~VI.

\section{Modeling City-sizes using the Rank-Order distribution }

\subsection{The Rank-Ordering (RO) distribution}
\label{SEC:ROD}

Let us first recapitulate the mathematical formulation of rank-ordered distribution, in particular the DGB distribution.  
Suppose we have data on the sizes of $N$ items (e.g., cities) arranged in decreasing order of ``importance" (e.g., size)
so that the $i$-th item (or city) has size $n_i$ with rank $r_i$.
Note that $(r_1, \ldots, r_N)$ is a permutation of $(1, \ldots, N)$.
The most widely used hyperbolic Pareto (Zipf's) law fits the data through the probability of rank $r$ being given by
\begin{equation}
f_P(r) = A\cdot\frac{1}{r^\nu}, ~~~~r=1, \ldots, N,
\label{EQ:Pareto}
\end{equation}  
where $\nu>0$ is a model parameter, known as the exponent, and  
$A$ is the normalizing constant defined by the formula 
$ A=\left[ \sum_{r=1}^N r^{-\nu}\right]^{-1}$.
As already noted in Section \ref{SEC:intro}, this Pareto or power law provides good fit to the data 
only at  low-ranks (large sizes).
This is mostly because of the fact that the empirical data distribution often have an inflection point, 
which is not captured by  the single power law (\ref{EQ:Pareto}). 

In order to accommodate the inflection as well as the shape of the two ends of the hierarchy, 
one can use the simple two-parameter RO distribution, namely the DGBD \cite{Martinez-Mekler09}, defined by 
\begin{equation}
f_{(a,b)}(r) = A\frac{(N+1-r)^b}{r^a}, ~~~~r=1, \ldots, N,
\label{EQ:RO}
\end{equation}  
where $a, b$ are two real valued model parameters and  $A$ is again the normalizing constant. 
Given $N$ and $(a,b)$, one can compute $A$ as 
\begin{equation}
A=\left[ \sum_{r=1}^N \frac {(N+1-r)^b}{r^a}\right]^{-1},
\label{EQ:RO_A}
\end{equation}
so that $f_{(a,b)}(\cdot) $ is a proper probability distribution, i.e. $\sum_{r=1}^N f_{(a,b)}(r) =1$.

\begin{figure}[!h]
	\centering
	\subfloat[\large $b=0$]{
		\includegraphics[width=0.34\textwidth]{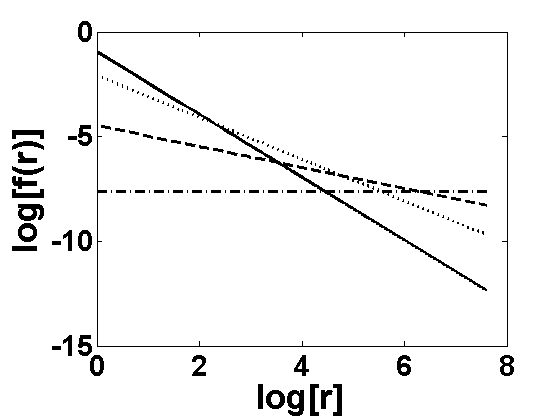}
		\label{FIG:b0ap}}
	~ 
	\subfloat[\large $b=0.5$]{
		\includegraphics[width=0.34\textwidth]{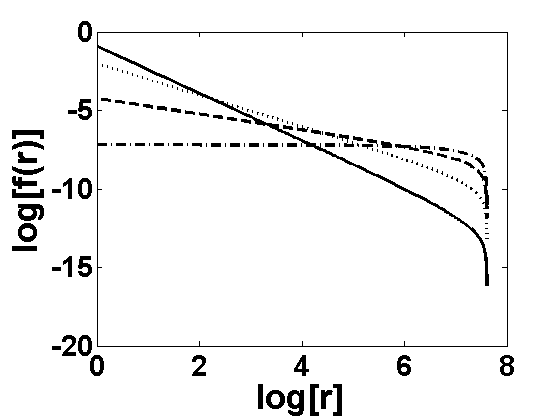}
		\label{FIG:loglogCumHaz_SmallCell}}
	\\
\subfloat[\large $b=0.3$]{
	\includegraphics[width=0.34\textwidth]{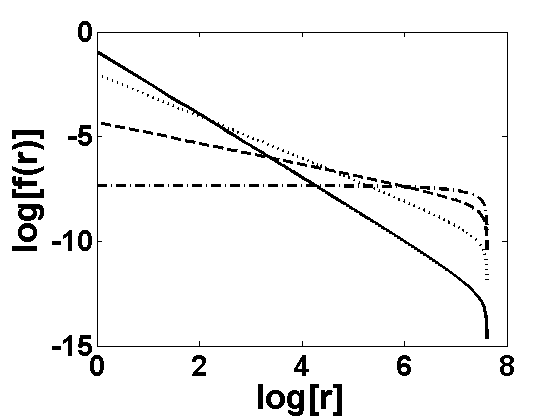}
	\label{FIG:loglogCumHaz_Veteran}}
~ 
\subfloat[\large $b=-0.3$]{
	\includegraphics[width=0.34\textwidth]{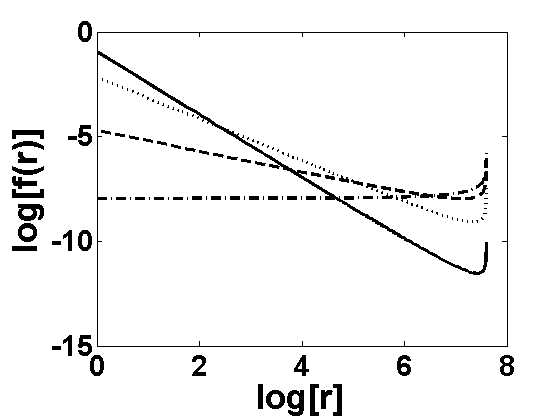}
	\label{FIG:RO_negB}}
\\	
\subfloat[\large $b=1$]{
	\includegraphics[width=0.34\textwidth]{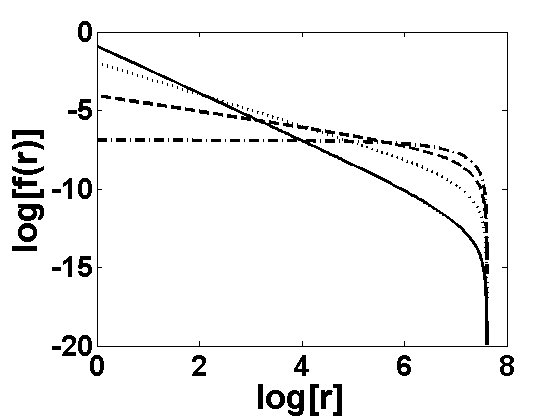}
	\label{FIG:loglogCumHaz_Veteran}}
~ 
\subfloat[\large $b=-1$]{
	\includegraphics[width=0.34\textwidth]{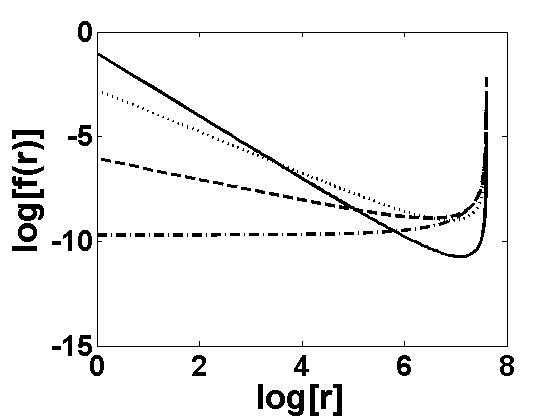}
	\label{FIG:loglogCumHaz_SmallCell}}
	\caption{The form of the DGBD $f\equiv f_{(a,b)}$, in log-log scale,  
		for different parameter values with $a\geq 0$ and $N=2000$
	[Dash-dotted line: $a=0$, Dashed: $a=0.5$, Dotted: $a=1$, Solid: $a=1.5$].}
	\label{FIG:RO_posA}
\end{figure}

The two parameters $a$ and $b$ indeed characterize the shape of the rank-size distribution 
at two-ends of low and high ranks, respectively, in respect to the inflection point.
Figure \ref{FIG:RO_posA}, representing  the shape of the DGBD for different values of $a, b\in\mathbb{R}$
 and $N=2000$ in the log-log scale,
clearly indicates its potentiality in modeling a wide variety of rank-size data structures;
see also \cite{Ausloos/Cerqueti:2016}.
In particular,  it is interesting to note that the case $b=0$ yields back 
the Zipf's law in (\ref{EQ:Pareto}) with $\nu = a$ having linear shape in the log-log scale 
(Figures \ref{FIG:b0ap});
these special cases indicate that there is no inflection point in the empirical data 
so that only one parameter ($a=\nu$) is sufficient  to model the slope of the data (in log-log scale).
Further, the shape of the DGBD becomes more flat as  $a$ and $b$ decreases close to zero,
just like the effect of $\nu$ on the power law. 
For any fixed $a$, the parameter $b$ in the DGBD allows a wide variety of different distributional graphs;
furthermore, for some exceptional data a non-monotonic DGBD may overall be the best fit although the data is monotonic.
This will be commented later on.

In terms of physical interpretation, the range of the distribution becomes more broad 
(spread over a larger interval and is less uniform) with increasing values of $a,b$, 
leading to easier discrimination between ranks in the opposite extremes. 
On the other hand, as the values of either $a$ or $b$ decreases, 
the sizes of the individual ranks become more uniform (lesser spread)
and hence more difficult is their discrimination;
the case $a=0=b$ coincides with the uniform distribution which is most uncertain and less broad.
This phenomenon can be well connected and characterized by the concept of entropy,
as detailed in the following subsection.

\subsection{Shannon Entropy of the distribution }  

A city is not and cannot in any sense be treated as an isolated system, since it always exchanges
materials, energy, information and people with its surroundings. 
When related to a probability, the concept of the Shannon-Gibb's entropy can be used in analyzing
the uncertainty in a city-size distribution.
For the DGBD, the Shannon entropy  \cite{Ausloos/Cerqueti:2016} is given by,
\begin{eqnarray}\label{entropy1}
S (a,b) &=& -\sum_{r=1}^N f_{(a,b)}(r) \log f_{(a,b)}(r)
= -\log A 
-A\sum_{r=1}^N \frac{(N+1-R)^b}{r^a} [b \log (N+1-r)-a \log r].
\end{eqnarray}
Note that, since $f_{(a,b)}(r) $ is assumed to be normalized to $1$ function, following \cite{Ausloos/Cerqueti:2016},
it is absolutely meaningful to use the Shannon entropy as defined in (\ref{entropy1}).   
The entropy measure in Eq.~(\ref{entropy1}) lies at the basis of information theory \cite{jaynes} and 
is applied successfully across several disciplines of arts and sciences \cite{kapurbook}.

\begin{figure}[!b]
	\centering
	\subfloat[$N=2000$]{
		\includegraphics[width=0.45\textwidth]{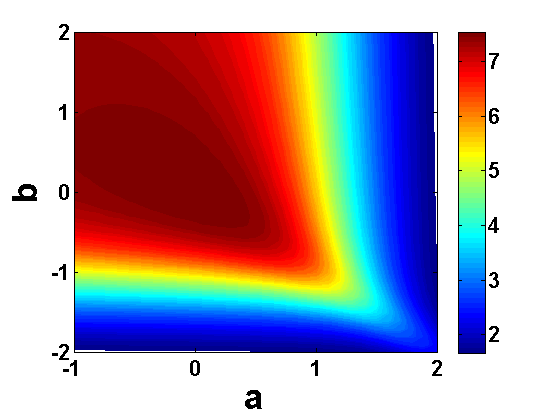}
		\label{FIG:loglogCumHaz_Veteran}}
	\subfloat[$N=200,000$]{
		\includegraphics[width=0.45\textwidth]{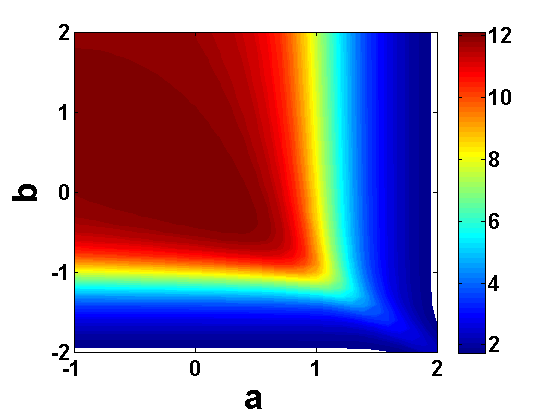}
		\label{FIG:loglogCumHaz_SmallCell}}
\\	
	\subfloat[Plot over varying $N$]{
	\includegraphics[width=0.7\textwidth]{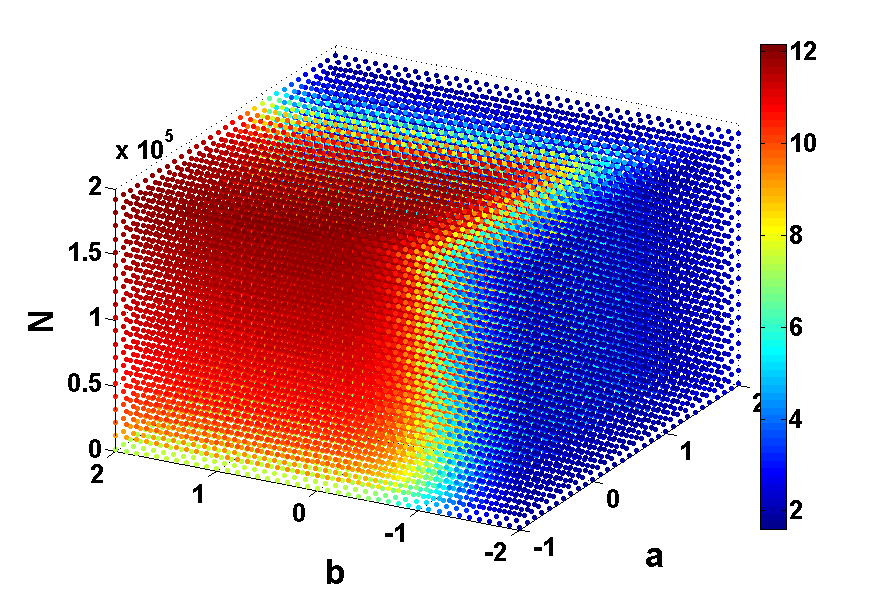}
	\label{FIG:loglogCumHaz_SmallCell}}
	\caption{Entropy of the DGBD over $(a, b)$ for different $N$}
	\label{FIG:Entropy_N20}
\end{figure}

Figures \ref{FIG:Entropy_N20} (a) and (b)  respectively display the contour plots for the exact values of the entropy corresponding to the DGBD with $N=2000$ and $200,000$ over the parameters $(a,b)$. 
Also, increasing one dimension in the plot, in Fig. 2(c) we have represented the entropy for different values of $N$ along with the variation of $(a,b)$. 
Clearly, the maximum entropy corresponding  to the case $a=b=0$  leads to the uniform distribution. 
The case $b=0$, with any $a >0 $, corresponds to the entropy for the power law distribution. 
Further, the pattern of the entropy is almost indifferent for different $N$ although the absolute value of $S$ changes. 
Hence, the city-size distributions can be compared in terms of their entropy and distributional structure 
only through the values of the two parameters $a,$ and $b$;
we will illustrate it further in Section \ref{SEC:discrimination}.

\subsection{Parameter Estimation through Maximum Likelihood}
\label{SEC:est}

We now encapsulate how the DGB distribution can be used for universal modeling of the city-size distributions 
and explain the methodology for the estimation of the parameters $(a,b)$ from the empirical data.
Suppose we have the data for a population of $N$ cities in order of decreasing sizes 
and $x_i$, $r_i$ respectively denote the actual size and rank of the $i$-th city for $i=1, \ldots, N$.
We can fit the DGB distribution to these data by means of estimating the parameters $(a, b)$ 
which leads the assumed 
 probabilities $f_{(a,b)} (r)$  to be $closest$ to the empirical (normalized) sizes in an appropriate sense.
The best (asymptotically) statistical estimation technique is the maximum likelihood estimator (MLE)
which measures the closeness in terms of Kullback-Leibler divergence measure
and maximizes the probability of the observed (sample) data under the assumed model. 
Since the Kullback-Leibler divergence is directly linked with the Shannon entropy 
$S$ in (\ref{entropy1}) and the corresponding cross-entropy,
the MLE is also the same as the corresponding minimum cross-entropy estimator, 
popular in information sciences and statistical physics.
The MLE also has minimum standard error in estimation, asymptotically for larger sample sizes, 
among a wide class of useful estimators.

Given the ranked city-size data $(x_i, r_i)$, the MLE of $(a, b)$ in the best fitted RO model  (DGBD) 
is to be obtained by maximizing the likelihood function given by 
\begin{equation}
L(a, b) = \prod_{i=1}^N f_{(a,b)}(r_i)^{{x}_i} = \prod_{i=1}^N \frac{(N+1-r_i)^{b{x}_i}}{r_i^{a{x}_i}}A^{{x}_i}.
\end{equation}
Equivalently, one can also maximize the log-likelihood function given by
\begin{eqnarray}
&&\ell(a, b)  = \log L(a, b) 
= b \sum_{i=1}^N {{x}_i}\log(N+1-r_i)-
a\sum_{i=1}^n{{x}_i}\log(r_i)+ \log(A)\sum_{i=1}^n{{x}_i},
\nonumber
\end{eqnarray} 
or solve the estimating equations 
$$\frac{\partial}{\partial a}\ell(a, b) = 0 = \frac{\partial}{\partial b}\ell(a, b).$$
Noting from Eq.~(\ref{EQ:RO}) that $A$ is a function of $(a, b)$, 
we can numerically maximize $\ell(a, b)$ or solve the above estimating equations, with respect to $a, b\in \mathbb{R}$,
to obtain their MLE, which we will denote as $(\widehat{a}, \widehat{b})$.  In all our data analyses we have numerically minimized the negative of log likelihood function $\ell (a,b)$, in the software MATLAB (version14a)
using the in-built function \textit{'fminsearch'}.

\subsection{Prediction and Goodness-of-fit}
\label{SEC:goodness}

Once we have obtained the MLE $(\widehat{a}, \widehat{b})$ of the parameters $(a, b)$ of 
the DGBD fitted to a given set of empirical data, we can predict the rank-sizes from this fitted model.
The predicted normalized size of rank $r_i$ is given by $f_{(\widehat{a}, \widehat{b})}(r_i)$
and hence, multiplying it by the total size $\sum\limits_{i=1}^N x_i$, one gets the predicted size of $r_i$
that can be denoted as 
$$p_i = \left(\sum_{i=1}^N x_i\right)f_{(\widehat{a}, \widehat{b})}(r_i).$$

If the estimated RO model is a good fit to the given data, the predicted values $p_i$ 
should be close to the corresponding observed size $x_i$ and the corresponding plot gives us a visual indication of the goodness-of-fit of our RO model (via DGBD) to any dataset.
A quantitative measure of fit can also be obtained by summing  the error in prediction;
we will here consider the measure defined in terms of the Kolmogorov-Smirnov distance between 
the predicted and observed cumulative frequencies of ranks as given by
\begin{eqnarray}
KS = \max_{1\leq i \leq N} \left|\left(\sum_{j: r_j\leq r_i} p_j\right) - \left(\sum_{j: r_j\leq r_i} x_j\right)\right|. 
\label{EQ:KS}
\end{eqnarray}
The use of cumulative sizes instead of actual sizes in the definition of KS 
provides better stability of this goodness-of-fit measure. 
Note that, this KS measure becomes zero if and only if $p_i=x_i$ for all $i=1, \ldots, N$,
i.e., all the predicted sizes coincide exactly with the observed sizes.
For a given dataset, the KS measures closer to zero indicate better goodness-of-fit 
and the model corresponding to minimum KS will be the $best$ as a predictive model of the underlying rank-size distribution.

\section{Data Description}
\label{SEC:data}

In the last few decades the human population is increasingly clustered in urban areas.  The urban population of the world has grown rapidly from 746 million in 1950 to 3.9 billion in 2014. Asia, despite its lower level of urbanization, is home to 53\% of the world's urban population, 
followed by Europe with 14\% and Latin America and the Caribbean with 13\% \cite{desa}.
Today, 55\% of the world's population lives in urban areas, a proportion that is expected to increase to 68\% by 2050. 
Projections show that urbanization, the gradual shift in residence of the human population from rural to urban areas, 
combined with the overall growth of the world's population could add another 2.5 billion people to urban areas by 2050, 
with close to 90\% of this increase taking place in Asia and Africa, according to a new United Nations data set launched in May, 2018 \cite{un}. 
Together, India, China and Nigeria will account for 35\% of the projected growth of the world's urban population between 2018 and 2050. 
By 2050, it is projected that India will have added 416 million urban dwellers, China 255 million and Nigeria 189 million.
In the present day context these data motivate us  to study the size-distribution of  {\it{all}} cities, or urban settlements
in different countries, in particular India and China, once again.

As an effect of urbanization,  it is now difficult to distinguish  city, suburb or town. All definitions of $city$ has shortcomings, either it is defined as incorporate area, an urban agglomeration, or a human settlement with population density larger than some threshold value \cite{plos2007}.
For our analysis,  we use the term $city$ to mean any human settlement that is denser than its surroundings; 
we consider towns, cities, counties, urban agglomerations according to the availability of the data for various countries
as described below;  these datasets are public and freely available. 
\begin{itemize}
\item Indian city-size data from the censuses at the years 1991, 2001 and 2011 \cite{india data}, 
with minimum inhabitant of 5,000.
\item Chinese city and county sizes for three years, namely 1990, 2000 and 2010, as defined in \cite{china data}.
\item USA data for each years between 2010 to 2017, with minimum size of 5,000,
where the city populations are estimated based on 2010 census \cite{usa data}.
\item Brazil city populations  for the years 1991, 2000, 2010 and 2017 (estimated), 
with a minimum of 20,000 inhabitants \cite{brazilcity}. 
\item City sizes of three African Countries, namely Uganda, Sudan and Algeria for the years 2014, 2008 and 2008, respectively, 
with minimum sizes of 15,000, 20,000 and 13,000 \cite{africa}.
\item Italian cities for the years 1981, 991, 2001, 2011 and 2017 (estimated) \cite{italy}, with minimum population of 50,000.
\item Sweden city size estimates for each fifth year from 1990 to 2015 and 2017 \cite{sweden}, 
having population greater than 10,000. 
\item Switzerland city-sizes for the years 1980, 1990, 2000 and 
the corresponding estimates for 2010 and 2017 \cite{switzerland},
with minimum inhabitants of 10,000.
\item Australian urban agglomerations (UA) (all) for the years 2011 and 2016 \cite{australia}.
\item We also have considered the population data for major cities around the world in the year 2018,
as obtained from \cite{world2018}. 
\item Finally, we have also tested our model with the data on the population ($> 5000 $) of 226 countries 
and dependent territories across the world, as defined and detailed in \cite{world2018}.
\end{itemize}

\section{Empirical Illustrations: Universality in space and time }

In this section, we present the results obtained by the applications of the DGBD 
for different countries' city-size data described in the previous section. 
For comparison, the popular Pareto model, corresponding to the Zipf's law, has also been fitted
to the same sets of data through a linear regression between logarithms of size and rank,
and the superiority of our proposal has been illustrated in all the cases through the KS goodness-of-fit measure (\ref{EQ:KS}). 
The highly significant fits of the DGBD for all such cases indicate its universality over time 
as well as its global nature to model different kinds of city size distributions 
from different economies and geographies  across the world.

\subsection{Size distribution for Indian cities}

Let us start our demonstration with Indian cities, 
using the census data for the years 1991, 2001 and 2011, and also remark on the viability of our prediction.  
According to the most recent census in 2011 \cite{census2011}, 
India is the second largest country in the world with more than 1.2 billion people, 
and it is predominantly  rural country with about 67\% population living in rural areas \cite{world2018}. 
The country has a total of 39 cities with more than one million residents each; 
among them, Mumbai and Delhi have populations exceeding 10 million. 
However, the country also has smaller but still very populated cities, 
including 388 having populations over 100,000,  and a whopping 2,483 cities with populations over 10,000.

\begin{figure}[!h]
	\centering
	\subfloat[Class I-V cities]{
		\includegraphics[width=0.5\textwidth]{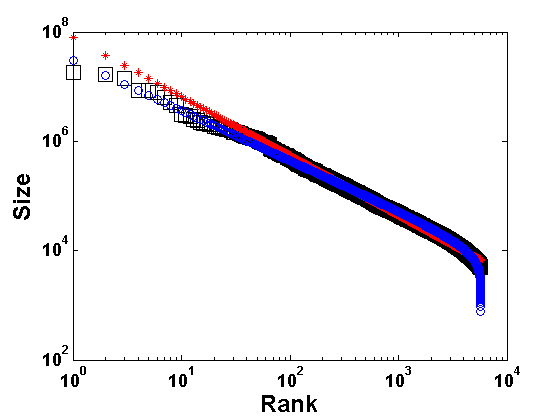}
		\label{FIG:loglogCumHaz_Veteran}}
	~ 
	\subfloat[Class I-IV cities]{
		\includegraphics[width=0.5\textwidth]{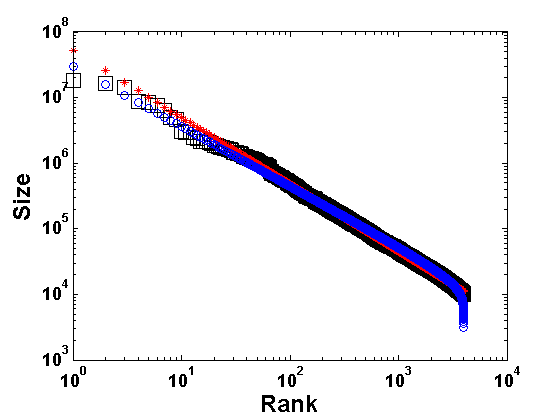}
		\label{FIG:loglogCumHaz_SmallCell}}
	\\
	\subfloat[Class I-II cities]{
		\includegraphics[width=0.5\textwidth]{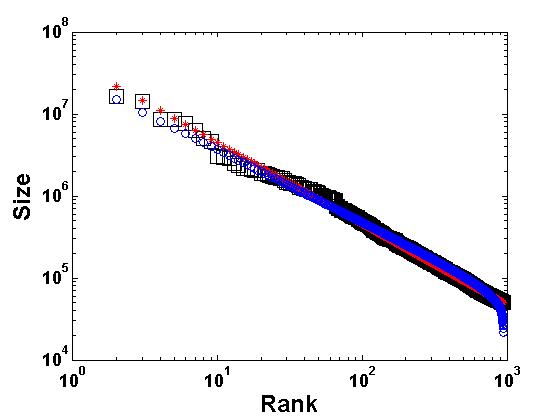}
		\label{FIG:loglogCumHaz_Veteran}}
	~ 
	\subfloat[Class I cities]{
		\includegraphics[width=0.5\textwidth]{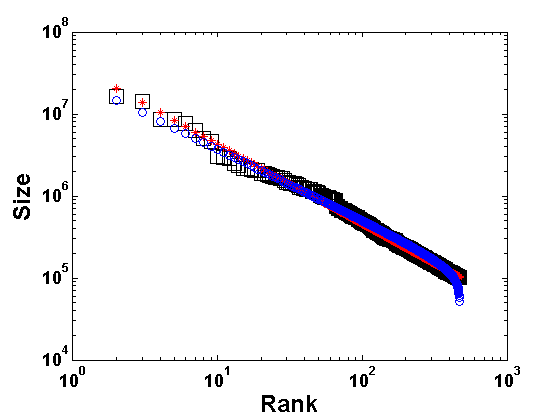}
		\label{FIG:loglogCumHaz_SmallCell}}
	\caption{Plots of the actual and the predicted sizes over ranks for different sets of Indian cities in 2011 
		(Black square: observed sizes, blue circle: RO fits, red star: Pareto fits)}
	\label{FIG:Single_2011}
\end{figure}

As noted earlier in Section \ref{SEC:data}, we  consider Indian cities to be defined 
as human settlements with  population more than  5,000. 
The whole data is divided into different  classes (I-V) according to its population
as follows: Class (I) is above 100,000 population, Class (II) is from 50,001 to 100,000,
Class (III) is from 20,001  to 50,000, Class (IV) is from 10,001 to 20,000   and Class (V) is from 5,000 to 10,000. 
We have fitted the DGBD for all these city-sizes (Class I-V) and 
also for some subset of cities starting with a larger minimum size.
In particular, we consider the subset of cities with population at least 100,000 (i.e., Class I cities),
or at least 50,000 (i.e., Class I and II cities) or at least 10,000 (i.e., Class I-IV cities).
The estimated parameter values and the goodness-of-fit measures (KS) for all four cases 
and three years are reported in  Table \ref{TAB:Single_2001}, along with total number of cities ($N$)
and the sizes $x_{\min}, x_{\max}$ of the smallest and the largest cities, respectively, in each cases.
The corresponding fitted (and actual) sizes are plotted in Figure \ref{FIG:Single_2011} for the most recent year 2011;
the plots for other two years are very similar and  not presented here to avoid  repetition.
The fitted values and goodness-of-fit measures corresponding  to the Pareto modeling (along with the estimated exponent $\nu$),
for the same sets of data, are also shown in Figure \ref{FIG:Single_2011} and Table \ref{TAB:Single_2001}, respectively.

\begin{table}[h]
	\caption{Parameter estimates for Indian Cities in different years, along with total number of cities ($N$) \\
		and the sizes $x_{\min}, x_{\max}$ of the smallest and the largest cities, respectively.}
	\centering	
	\begin{tabular}{l|rrr|rrr|rr}\hline
		City & \multicolumn{3}{c}{Summary}	& \multicolumn{3}{|c|}{RO Fit} & \multicolumn{2}{c}{Pareto Fit} \\
		Classes	&	$N$	& $x_{\min}$ & $x_{\max}$ &	$\widehat{a}$	&	$\widehat{b}$	&	KS	&	$\nu$	&	KS	\\\hline
\multicolumn{9}{l}{2011}\\\hline
I		&	468	& 100039	& 18394912	&	0.8552	&	0.1613	&	0.0308	&	0.9684	&	0.1252	\\
I-II	&	942	&	50087 &	18394912 &		0.8791	&	0.1641	&	0.0313	&	0.9890	&	0.1470	\\
I-IV	&	4001	& 10001	&	18394912	&	0.9122	&	0.1899	&	0.0310	&	1.0190	&	0.1923	\\
I-V		&	5749	&	5001 &	18394912	&	0.9144	&	0.3051	&	0.0304	&	1.0856	&	0.4418	\\\hline
\multicolumn{9}{l}{2001}\\\hline
I		&	394		& 100065	&	16434386 &		0.8842	&	0.0671	&	0.0275	&	0.9328	&	0.0506	\\
I-II	&	798		& 50057	&	16434386 &		0.8940	&	0.0610	&	0.0256	&	0.9333	&	0.0462	\\
I-IV	&	3307	& 10005 &	16434386	&	0.9079	&	0.1349	&	0.0229	&	0.9923	&	0.1560	\\
I-V		&	4186	& 5002  &	16434386	&	0.9056	&	0.2411	&	0.0220	&	1.0535	&	0.3605	\\
		\hline
\multicolumn{9}{l}{1991}	\\\hline
I		&	289		& 100235	&	9925891 &		0.8072	&	0.1374	&	0.0204	&	0.9129	&	0.0912	\\
I-II	&	614		& 50048	&	9925891	&	0.8347	&	0.1166	&	0.0233	&	0.9182	&	0.0838	\\
I-IV	&	2533	& 10005	&	9925891	&	0.8638	&	0.1823	&	0.0226	&	0.976	&	0.1731	\\
I-V		&	3138	& 5004	&	9925891	&	0.8614	&	0.2964	&	0.0216	&	1.038	&	0.3642	\\
\hline
\end{tabular}
	\label{TAB:Single_2001}
	
\end{table}

\vspace{1in}	
The superiority of the goodness-of-fit in the proposed RO distribution, the DGBD, over the usual Pareto law
is clearly visible from Figure \ref{FIG:Single_2011} and Table \ref{TAB:Single_2001} in all cases.
In particular, when we consider a wider range of cities having sizes significantly small as well as significantly large 
(i.e., enough observations and variety in both the high and low end of the ranks leading to larger $N$),
the Pareto model fails remarkably as also noted earlier 
\cite{12kgbb, 13kgbb, 14kgbb, 15kgbb, 16kgbb, 17kgbb, 18kgbb, 19kgbb,luck17I,luck17U}.
In such cases, our proposal through rank ordering can provide excellent fit with very little prediction error (KS) 
that is almost the same as that obtained with only large sizes (low-ends of the rank) of cities.
This observation, along with establishing the significance of the proposed approach over the existing Zipf's law
of city size distributions, also indicate the universality of the RO distribution 
in predicting the city distributions with wider ranges (in both extremes) equally well.


\subsection{Chinese City-size Distribution}

\begin{figure}[!th]
	\centering
	\subfloat[1990, City]{
		\includegraphics[width=0.33\textwidth]{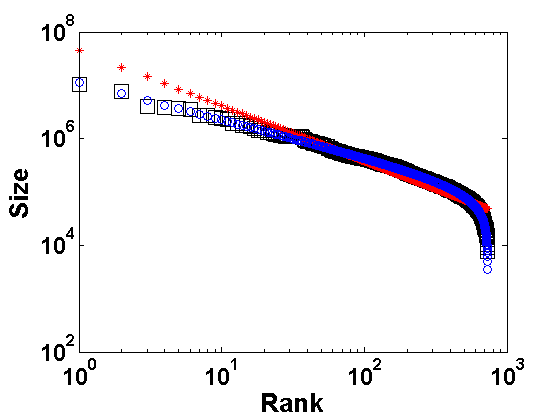}
		\label{FIG:loglogCumHaz_Veteran}}
	~ 
	\subfloat[2000, City]{
		\includegraphics[width=0.33\textwidth]{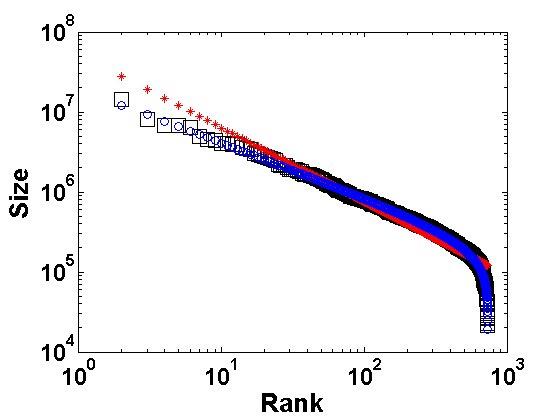}
		\label{FIG:loglogCumHaz_SmallCell}}
	~ 
	\subfloat[2010, City]{
		\includegraphics[width=0.33\textwidth]{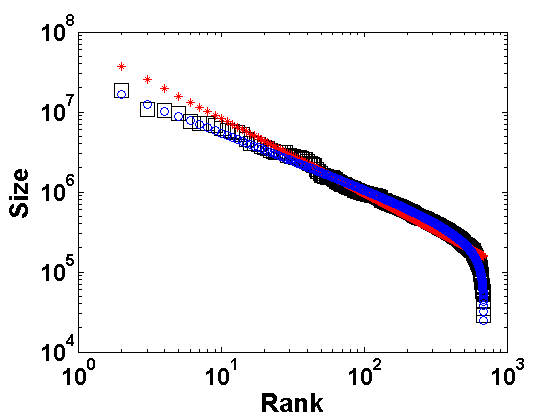}
		\label{FIG:loglogCumHaz_SmallCell}}
	\\
	\subfloat[1990, County]{
		\includegraphics[width=0.33\textwidth]{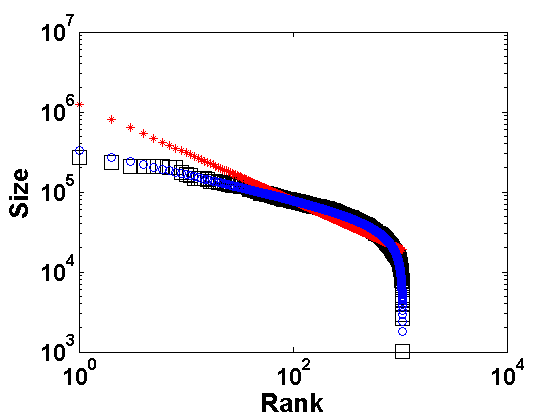}
		\label{FIG:loglogCumHaz_Veteran}}
	~ 
	\subfloat[2000, County]{
		\includegraphics[width=0.33\textwidth]{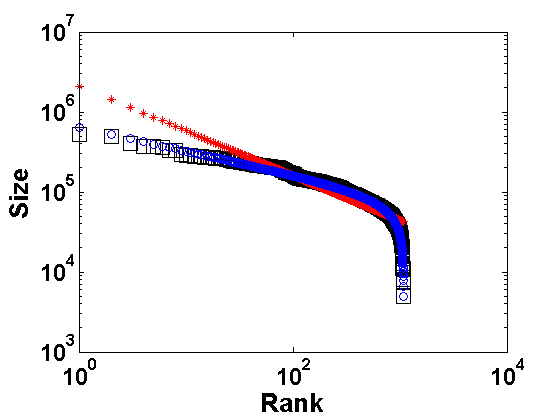}
		\label{FIG:loglogCumHaz_SmallCell}}
	~ 
	\subfloat[2010, County]{
		\includegraphics[width=0.33\textwidth]{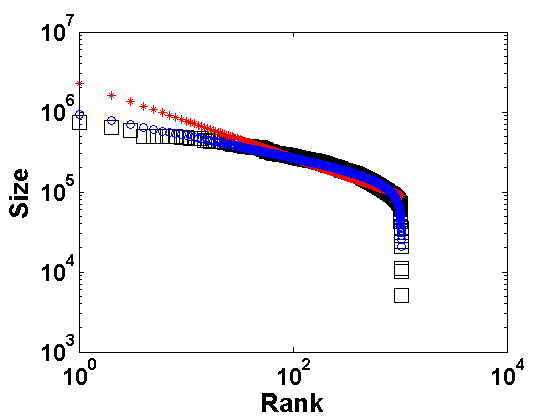}
		\label{FIG:loglogCumHaz_SmallCell}}
	\\
	\subfloat[1990, City+County]{
		\includegraphics[width=0.33\textwidth]{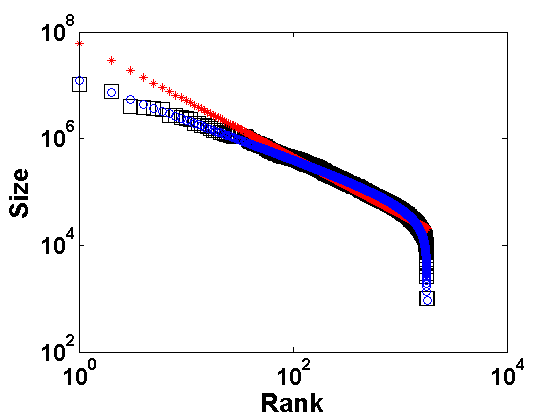}
		\label{FIG:loglogCumHaz_Veteran}}
	~ 
	\subfloat[2000, City+County]{
		\includegraphics[width=0.33\textwidth]{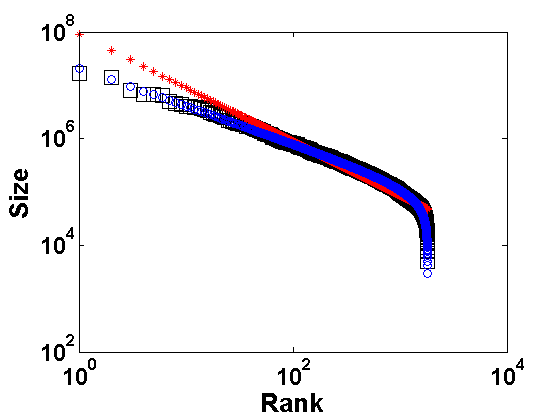}
		\label{FIG:loglogCumHaz_SmallCell}}
	~ 
	\subfloat[2010, City+County]{
		\includegraphics[width=0.33\textwidth]{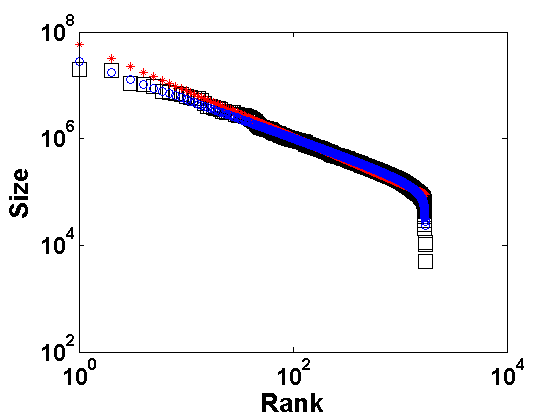}
		\label{FIG:loglogCumHaz_SmallCell}}
	\caption{Plots of the actual and the predicted sizes over ranks for Chinese data
	(Black square: observed sizes, blue circle: RO fits, red star: Pareto fits)}
	\label{FIG:China}
\end{figure}

China has an  importance  in urban planing and city-size analysis 
as it is the most populous country in the world with  1,415,045,928 people according to a 2018 estimate \cite{world2018}. 
According to  \cite{world2018}, Shanghai is the most populous city with an estimate of 25,582, 138 in 2018. 
We consider the sizes of all Chinese \textit{Cities} and \textit{Counties},
as defined in \cite{china data},  for the years 1990, 2000 and 2010.
Both the DGBD and the usual Pareto distribution is fitted for the sizes of these cities and counties separately 
as well as to their combined pool. 
The fitted sizes obtained in both the approaches are plotted in Figure \ref{FIG:China},
whereas the estimated parameters and KS measures are reported in Table \ref{TAB:China}.

\begin{table}[h]
	\caption{Estimated measures for Chinese Cities, , along with total number of cities ($N$)
		and \\ the sizes $x_{\min}, x_{\max}$ of the smallest and the largest cities, respectively.}
	\centering	
	\begin{tabular}{l|rrr|rrr|rr}\hline
	 & \multicolumn{3}{c}{Summary}	& \multicolumn{3}{|c|}{RO Fit} & \multicolumn{2}{c}{Pareto Fit} \\
	Year	&	$N$	& $x_{\min}$ & $x_{\max}$ &	$\widehat{a}$	&	$\widehat{b}$	&	KS	&	$\nu$	&	KS	\\\hline
		\multicolumn{9}{l}{City Only}\\\hline
		1990	&	731	& 7761 	& 10429722 	& 	0.6939	&	0.5314	&	0.0081	&	1.037	&	0.4959	\\
		2000	&	735	& 21293 & 16704306 	&	0.6786	&	0.3695	&	0.0106	&	0.9212	&	0.2665	\\
		2010	&	687	& 28783 & 20217748 &	0.6899	&	0.3786	&	0.0207	&	0.9346	&	0.2632	\\
		\hline
		\multicolumn{9}{l}{County Only}\\\hline
		1990	&	1057 & 1006 & 266907 &	0.3073	&	0.4435	&	0.007	&	0.6005	&	0.1303	\\
		2000	&	1074 & 4987 & 522369 & 0.2932	&	0.4016	&	0.0104	&	0.5603	&	0.1033	\\
		2010	&	1028 & 5070	& 730810 &	0.2685	&	0.2801	&	0.0132	&	0.4627	&	0.0622	\\
		\hline
		\multicolumn{9}{l}{City + County Combined}\\\hline
		1990	&	1788 	& 1006 	& 10429722 	& 	0.7382	&	0.5271	&	0.0188	&	1.0637	&	0.6371	\\
		2000	&	1809 	& 4987	& 16704306 	&	0.7086	&	0.47	&	0.0106	&	1.0061	&	0.5065	\\
		2010	&	1715	& 5070	& 20217748 	&	0.7176	&	0.2333	&	0.0159	&	0.8675	&	0.1737	\\
		\hline
	\end{tabular}
	\label{TAB:China}
\end{table}

 One can note that, in all the cases the DGBD gives much improved fit to the city sizes
 compared to the Zipfs law. This is in-line with the findings of \cite{kgbb} 
 where the authors have shown that the Pareto modeling works well for Chinese data 
 only if we consider the restricted sets of cities having sizes greater than 150000 or 200000 in 1990 and 2000, respectively. 
 On the other hand, the proposed RO modeling fits the whole data of all cities or counties 
 (as well as their combined pool) very accurately leading to a very small error value (KS).
 Figure \ref{FIG:China} clearly demonstrates that, in all the cases,  
 the fits of the DGBD are also much better compared to the Pareto distribution in the observed data.

\subsection{Yearly City-size Distribution of USA}

\begin{figure}[!h]
	\centering
	\subfloat[2010]{
		\includegraphics[width=0.47\textwidth]{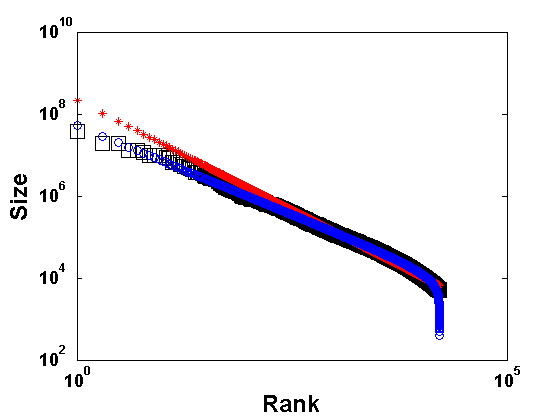}
		\label{FIG:loglogCumHaz_Veteran}}
	~ 
	\subfloat[2017]{
		\includegraphics[width=0.47\textwidth]{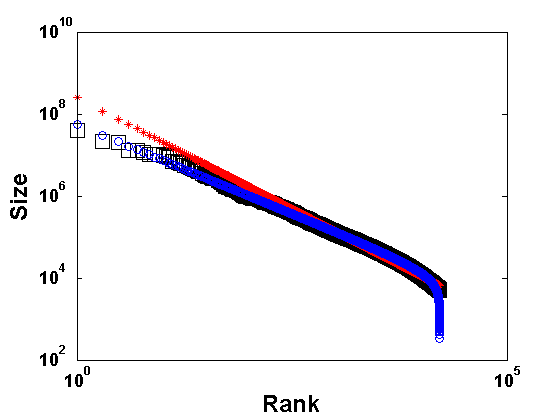}
		\label{FIG:loglogCumHaz_SmallCell}}
	\caption{Plot of the actual and the predicted sizes over ranks for USA city sizes (estimated) in 2010 and 2017
	(Black square: observed sizes, blue circle: RO fits, red star: Pareto fits)}
	\label{FIG:USA}
\end{figure}

USA is the 3rd most  populous country in the world according to the 2018 ranking \cite{world2018}. 
New York City, is the most populous city in USA (world city ranking 2) 
with a present population of 8,550,405 and expected to reach 9 million by 2040. 
However, USA has a developed economy in contrast to the developing ones in India and China; one may expect that USA city size distribution will be different from those of India and China if the cities with lower population are included in the study of city size distribution. 
To illustrate the universality and global nature of the DGBD
in modeling city sizes of any kind of economy, let us now apply our proposal for USA city size data (estimated),
as already described  in Section \ref{SEC:data}.
We consider only those entries having size greater than or equal 5,000 as a city.
The parameter estimates and goodness-of-fit measures (KS) obtained by 
both the Pareto and proposed RO modeling are given in Table \ref{TAB:USA} for each years from 2010 to 2017;
results for 2010 are obtained both from the true census data as well as the estimated data for comparison purpose. 
However, due to the resemblance of all these results, 
only the plots of the corresponding fitted sizes for the years 2010 and 2017 (both with estimated data ) are shown in Figure \ref{FIG:USA}. 

\begin{table}[h]
	\caption{Estimated measures for USA Cities, , along with total number of cities ($N$)
		and \\ the sizes $x_{\min}, x_{\max}$ of the smallest and the largest cities, respectively 
		(2010* denotes the census data; others are estimated data).}
	\centering	
	\begin{tabular}{l|rrr|rrr|rr}\hline
	& \multicolumn{3}{c}{Summary}	& \multicolumn{3}{|c|}{RO Fit} & \multicolumn{2}{c}{Pareto Fit} \\
	Year	&	$N$	&	 $x_{\min}$ & $x_{\max}$ &	$\widehat{a}$	&	$\widehat{b}$	&	KS	&	$\nu$	&	KS	\\\hline
2010*	&	16330	& 5000 & 37253956 &	0.8667	&	0.3474	&	0.0276	&	1.0741	&	0.7011	\\\hline
2010	&	16397	& 5001 & 37327690 &	0.8663	&	0.3466	&	0.0274	&	1.0736	&	0.7020	\\
2011	&	16412	& 5000 & 37672654 &	0.8665	&	0.3504	&	0.0275	&	1.0757	&	0.7128	\\
2012	&	16418	& 5000 & 38019006 &	0.8667	&	0.3544	&	0.0277	&	1.0778	&	0.7235	\\
2013	&	16443	& 5000 & 38347383 &	0.8669	&	0.3586	&	0.0278	&	1.0800	&	0.7359	\\
2014	&	16436	& 5000 & 38701278 &	0.8670	&	0.3614	&	0.0279	&	1.0816	&	0.7436	\\
2015	&	16449	& 5000 & 39032444 &	0.8672	&	0.3650	&	0.0280	&	1.0836	&	0.7545	\\
2016	&	16456	& 5001 & 39296476 &	0.8673	&	0.3685	&	0.0281	&	1.0853	&	0.7640	\\
2017	&	16459	& 5002 & 39536653 &	0.8673	&	0.3709	&	0.0282	&	1.0865	&	0.7708	\\
		\hline
	\end{tabular}
	\label{TAB:USA}
\end{table}

Once again our  RO approach fits the USA data much more accurately than the existing Pareto law. 
For each year, the accuracy obtained by the RO approach is extremely small and quite stable,
but the accuracy of Pareto modeling is about 25-30 times  lower 
and also increases over the year 
due to the increase in the variation of city sizes.

\subsection{City-size Distribution of Brazil}

We now consider another large developing country, namely Brazil, from South American continent.
In the population wise global ranking of 2018 estimate, 
Brazil is in 5th position with a population of 211,179,426 \cite{brazilcity}. 
We use the census data of the Brazilian city sizes as described in Section \ref{SEC:data}
and apply the RO modeling along with the usual Pareto modeling
for the cities having at least 20,000 inhabitants (as per the availability of data).
Resulting estimates are shown in Table \ref{TAB:Brazil}, 
which again show appreciably  improved fit by our proposal with the DGBD
as compared to the usual Pareto law.
The fitted values obtained by the DGBD 
is again seen to be much closer to the original city-sizes  compared to the power law fits. The  plots are not presented here for brevity.

\begin{table}[h]
	\caption{Estimated measures for Brazilian Cities with at least 20000 inhabitants, along with total number of cities ($N$)
		and \\ the sizes $x_{\min}, x_{\max}$ of the smallest and the largest cities, respectively.
		(* denotes estimated data)}
	\centering	
	\begin{tabular}{l|rrr|rrr|rr}\hline
	& \multicolumn{3}{c}{Summary}	& \multicolumn{3}{|c|}{RO Fit} & \multicolumn{2}{c}{Pareto Fit} \\
	Year	&	$N$	& $x_{\min}$ & $x_{\max}$ &	$\widehat{a}$	&	$\widehat{b}$	&	KS	&	$\nu$	&	KS	\\\hline
1991	&	709		& 20002 & 9412894 	&	0.886	&	0.079	&	0.021	&	0.947	&	0.096	\\
2000	&	922		& 20022 & 9813187 	&	0.845	&	0.157	&	0.012	&	0.962	&	0.176	\\
2010	&	1096	& 20002 & 11152344 	&	0.837	&	0.170	&	0.011	&	0.964	&	0.198	\\
2017*	&	1204	& 20030 & 11998090 	&	0.833	&	0.190	&	0.013	&	0.974	&	0.228	\\
		\hline
	\end{tabular}
	\label{TAB:Brazil}
\end{table}

\subsection{Distribution Cities in Three African Countries}

Next we try to test if the proposed approach fits well even for the African countries with underdeveloped economy.
In March 2013, Africa was identified as the world's poorest inhabited continent. 
Although, in 2017 the African Development Bank reported Africa to be the world's second-fastest growing economy, 
and estimated that average growth will rebound to 3.4\% in 2017, 
while the growth is expected to increase to 4.3\% in 2018 \cite{afr2017}. 
In this respect,  it is interesting to study the city size distribution of some  countries in Africa;
however, in most such countries, proper census data are rarely available. 
We here present the analysis for three such countries, namely Uganda, Algeria and Sudan,
where the latest census data have been found for the years 2014, 2008 and 2008, respectively, from \cite{africa}. 
Based on the availability of these data, we could only consider the cities having at least 15,000,
13,000 and 20,000 inhabitants, respectively, for the three countries. 
We would like to make a note here that, 
although Nigeria is going to become a good choice for the urban dwellers in coming years \cite{desa}, 
due to the unavailability of the data to the authors, 
the city-size distribution of Nigerian cities  could not  studied here.

Algeria is Africa's largest country and has several cities with population of over a million residents. 
The country's capital and largest city is Algiers with a population of around 3.5 million in 2008; 
the city of Oran is the second-largest having around 650,000 residents. 
Uganda is the 81st largest country by area in the world, but 36th in population. 
The largest city in Uganda is Kampala, with a population of 1,353,189 people in 2014. 
Uganda has one city with more than a million people, two cities with people between 100,000 to 1 million, 
but most cities (62) have smaller sizes between 10,000 to 100,000. 
Sudan has two cities with more than a million people, 19 cities of sizes between 100,000 to 1 million, 
and 42 cities of sizes between 10,000 to 100,000. 
The largest city in Sudan is Khartoum, with a population of 1,974,647 people in 2008.

Applying  both the RO and the Pareto modeling to the city-size data of these three countries, 
we report the resulting estimates and fitness measures in Table \ref{TAB:Africa}. 
For Uganda and Algeria, the proposed RO modeling using the DGBD performs as a close contender  of the Pareto law
which also gives reasonably well fit; note that the estimated $b$ parameter in the DGBD is close to zero
in both the cases indicating the good behavior of the Pareto modeling.
But, for Sudan, the estimate of $b$ is slightly away from zero and the resulting DGBD
provides a better fit with almost half the KS error as compared to the power law approach.

\begin{table}[h]
	\caption{Estimated measures for three African Countries, along with total number of cities ($N$)
		and \\ the sizes $x_{\min}, x_{\max}$ of the smallest and the largest cities, respectively.}
	\centering	
	\begin{tabular}{ll|rrr|rrr|rr}\hline
&	& \multicolumn{3}{c}{Summary}	& \multicolumn{3}{|c|}{RO Fit} & \multicolumn{2}{c}{Pareto Fit} \\
Country	& Year	&	$N$	& $x_{\min}$ & $x_{\max}$ &	$\widehat{a}$	&	$\widehat{b}$	&	KS	&	$\nu$	&	KS	\\\hline
Uganda & 2014	&	105	& 15056 & 1507080 &	0.917	&	0.029	&	0.048	&	0.936	&	0.046	\\
Algeria & 2008	&	180	& 13029 & 2364230 &	0.800	&	0.024	&	0.044	&	0.798	&	0.046	\\
Sudan & 2008	&	63	& 20302 & 1849659 &	1.033	&	0.187	&	0.043	&	1.157	&	0.094	\\
		\hline
	\end{tabular}
	\label{TAB:Africa}
\end{table}

\subsection{Distribution of Australian Agglomerations}

We next focus our attention on the 6th largest nation in the world, with very low population density; 
we test the RO model, through DGBD, to characterize the sizes of Urban Agglomeration (UA) of Australia. 
Australia has one of the most urbanized societies in the world.  
As of 2018, Australia has an estimated population of 24.77 million, up from the official 2011 census results of 21.5 million.
It is the most populous country in Oceania, three times more populous than its neighbor Papua New Guinea (8.2 million) 
and 5 times more populous than New Zealand (4.5 million), yet ranks only 54th in the world in terms of its population.

As an illustration, let us consider the data on the sizes of 101 UA in the years 2011 and 2016, obtained from \cite{australia}. 
When these Australian UA sizes are modeled by the DGBD and the Pareto distributions,
as before, the DGBD provides slightly improved fit compared to the Pareto fit;
see Table \ref{TAB:Australia}. All parameter estimates are surprisingly stable in both the year
indicating the stability in the urban dynamics of Australia even in the 5 year span.

\begin{table}[h]
	\caption{Estimated measures for Australian UA, along with total numbers ($N$)
		and their minimum and \\ maximum sizes $x_{\min}, x_{\max}$, respectively.}
	\centering	
	\begin{tabular}{l|rrr|rrr|rr}\hline
	& \multicolumn{3}{c}{Summary}	& \multicolumn{3}{|c|}{RO Fit} & \multicolumn{2}{c}{Pareto Fit} \\
	Year	&	$N$	& $x_{\min}$ & $x_{\max}$ &	$\widehat{a}$	&	$\widehat{b}$	&	KS	&	$\nu$	&	KS	\\\hline
		2011	&	101	& 6616  & 4034911 &	1.259	& 0.418	& 0.104	& 1.420	& 0.158\\
		2016	&	101	& 10288 & 4446807 &	1.267	& 0.434	& 0.106	& 1.434	& 0.163\\
		\hline
	\end{tabular}
	\label{TAB:Australia}
\end{table}

\subsection{City-size distributions of Three European Countries}
\label{SEC:Europe}

We now consider three smaller European countries, namely Italy, Sweden and Switzerland, all having strong developed economy.
The current population of Italy is 59,277,324, based on the latest United Nations estimates \cite{un},
which is equivalent to 0.78\% of the total world population (ranks 23rd). 
But, as large as 71.8\% of the population of Italy is urban (42,587,390 people in 2018 estimate);
the most populated city Roma (Rome) has population 2,648,843, followed by Milano (Milan)  with 1,305,591 \cite{italy}.  
In the Scandinavian country Sweden, almost all of the population live in urban areas;  
it has one city with more than a million people (Stockholm, with population 1,515,017 people), 
9 cities of sizes between 100,000 to 1 million, 
and as many as 141 small cities of population between 10,000 to 100,000 \cite{sweden}. 
Our third choice Switzerland, however,  has no city  with more than a million people, 
5 cities having 100,000 to 1 million people each, and 145 cities having 10,000 to 100,000 people;
the largest city Zurich has a population of only 341,730 residents \cite{switzerland}.

We repeat our analysis 
with  the census data of Italy for the years 1981, 1991, 2001 and 2011,  
and the corresponding estimated city sizes for the year 2017,
 restricting  our attention to the human settlements having at least 50,000 inhabitants.
We also apply the same for the yearly official estimate of city sizes of Sweden, 
from the year 1990 to 2015 and 2017, having at least 10,000 inhabitants.
For Switzerland we fit the distribution for the census data of the year 1980, 1990 and 2000,
and the official estimates for the year	2010 and 2017, 
where again we restrict only to the settlements having at least 10,000 human inhabitants. 
The resulting estimates and the KS measure of fits are provided in Table \ref{TAB:Italy} 
for all the three countries.

\begin{table}[h]
	\caption{Estimated measures for Cities of Italy, Sweden and Switzerland, along with total number of cities ($N$) \\ 
		and the sizes $x_{\min}, x_{\max}$ of the smallest and the largest cities, respectively.  
		(* denotes estimated data)}
	\centering	
	\begin{tabular}{l|rrr|rrr|rr}\hline
		& \multicolumn{3}{c}{Summary}	& \multicolumn{3}{|c|}{RO Fit} & \multicolumn{2}{c}{Pareto Fit} \\
		Year	&	$N$	& $x_{\min}$ & $x_{\max}$ &	$\widehat{a}$	&	$\widehat{b}$	&	KS	&	$\nu$	&	KS	\\\hline
		\multicolumn{9}{l}{Italy, having at least 50000 inhabitants}\\\hline
		1981	&	121	& 50666 & 2840259 &	0.898	&	$-$0.085	&	0.018	&	0.820	&	0.072	\\
		1991	&	128	& 50018 & 2775250 &	0.886	&	$-$0.122	&	0.017	&	0.777	&	0.086	\\
		2001	&	134	& 50032 & 2546804 &	0.865	&	$-$0.130	&	0.018	&	0.749	&	0.088	\\
		2011	&	141	& 50013 & 2617175 & 	0.857	&	$-$0.140	&	0.017	&	0.731	&	0.091	\\
		2017*	&	144	& 50645 & 2872800 &	0.872	&	$-$0.152	&	0.018	&	0.733	&	0.097	\\
		\hline
		\multicolumn{9}{l}{Sweden, having at least 10000 inhabitants}\\\hline
		1990*	&	107	& 10041 & 1209557 &	1.061	&	$-$0.202	&	0.052	&	0.863	&	0.160	\\
		1995*	&	111	& 10000 & 1275492 &	1.066	&	$-$0.204	&	0.051	&	0.870	&	0.158	\\
		2000*	&	110	& 10267 & 1344798 &	1.085	&	$-$0.213	&	0.050	&	0.884	&	0.162	\\
		2005*	&	112	& 10091 & 1388306 &	1.085	&	$-$0.204	&	0.050	&	0.892	&	0.159	\\
		2010*	&	117	& 10037 & 1517006 &	1.096	&	$-$0.209	&	0.049	&	0.901	&	0.161	\\
		2015*	&	123	& 10023 & 1654623 &	1.102	&	$-$0.210	&	0.051	&	0.908	&	0.163	\\
		2017*	&	125	& 10028 & 1705718 &	1.103	&	$-$0.209	&	0.051	&	0.912	&	0.162	\\\hline
		\multicolumn{9}{l}{Switzerland, having at least 10000 inhabitants}\\\hline
		1980	&	103	& 10001 & 369522 &	0.846	&	$-$0.128	&	0.022	&	0.729	&	0.093	\\
		1990	&	118	& 10180 & 365043 &	0.827	&	$-$0.142	&	0.021	&	0.698	&	0.096	\\
		2000	&	125	& 10142 & 363273 &	0.807	&	$-$0.138	&	0.018	&	0.680	&	0.090	\\
		2010*	&	143	& 10003 & 372857 &	0.778	&	$-$0.125	&	0.019	&	0.659	&	0.083	\\
		2017*	&	155	& 10007 & 409241 &	0.772	&	$-$0.117	&	0.020	&	0.659	&	0.081	\\
		\hline
	\end{tabular}
	\label{TAB:Italy}
\end{table}

 Although Table  \ref{TAB:Italy} illustrates   significantly improved goodness of the RO modeling over the usual Pareto law, but an interesting observation of negative $\hat{b}$ values, unlike the $\hat{b}$ values of other countries, entails  some explanation. 
Among various possible functional  forms of the RO distributional assumption (Figure \ref{FIG:RO_posA}), 
in case the rank is defined in decreasing order of importance (as in the case of city sizes here), 
it is intuitive to consider those RO distributions (in particular, those DGBDs) 
providing a monotonous non-increasing structure
which indeed restricts the parameter choice as $(a, b)\geq 0$. 
However, the DGBD with $a>0$ but very small negative value for $b$,
as obtained for these three European countries, indeed has an almost monotone structure
except for the extreme higher end (smaller sizes) where it  increases slightly; see Figure \ref{FIG:RO_negB}.
This suggests that  to get the best fit for the city size distributions of these 3 European countries  we are indeed loosing the monotonicity requirement while considering 
 the cities with 
 smaller  sizes.  The physical reasoning behind this  observation lies within  the population data; the  data exhibits that the extremely small cities of these 3 countries  are of almost equal sizes and it is difficult to individually 
 rank them ( only in the higher rank end).  
Note that, the major part of the city sizes including the low-rank end has a strictly monotone structure
allowing their easy discrimination from the smaller cities.  
So, accommodating the negative values of the parameters, we can very easily discriminate such countries having 
more uniform city size distributions at the lower end  from other countries 
having clearly distinct city sizes across its full spectrum 
(like India, China, USA, Brazil). 
On the contrary, if one wants to strictly retain 
 the monotone structure in the full spectrum of the city-size distribution, 
then the best fit for all three European countries is given by the power law (restricting $b$ to be non-negative). It may be noted that, unlike other countries studied above, the KS error in Pareto fit is reasonably small 
for Italy, Sweden and Switzerland.

\subsection{World-wide Size distributions of Cities and Countries in the year 2018} 

If we focus on the word-wide city size ranking, it is noted that at present, 
Tokyo is the world’s largest city with an agglomeration of 37 million inhabitants, 
followed by New Delhi with 29 million, Shanghai with 26 million, 
and Mexico City and Sao Paulo, each with around 22 million inhabitants \cite{un}. 
Today, Cairo, Mumbai, Beijing and Dhaka all have close to 20 million inhabitants. 
By 2020, Tokyo’s population is projected to begin to decline, 
while Delhi is projected to continue growing and to become the most populous city in the world around 2028.
By 2030, the world is projected to have 43 megacities with more than 10 million inhabitants, 
most of them in developing regions. 
However, some of the fastest-growing urban agglomerations are cities with fewer than 1 million inhabitants, 
many of them located in Asia and Africa. While one in eight people live in 33 megacities worldwide, 
close to half of the world’s urban dwellers reside in much smaller settlements with fewer than 500,000 inhabitants.
As the world continues to urbanize, sustainable development depends increasingly on 
the successful management of urban growth, especially in low-income and lower-middle-income countries 
where the pace of urbanization is projected to be the fastest. 
Understanding the key trends in urbanization likely to unfold over the coming years is very crucial 
and this motivated us to analyze the size distributions of the top 20 cities of 230 countries 
and dependent territories around the world for the year 2018; see  \cite{world2018} for their details. 
As before, we have restricted our modeling to the cities with a minimum number of 5000 habitants, 
but still the filtered data of 3371 cities have a very wider spectrum of population varying from 5017 to 37,468,302.
These are the estimated sizes projected from the actual census data collected in the last census years 
(which is different for different countries). 
The complete listing shows that,  although Tokyo tops the rank-wise list of top cities around the world, 
 there are many cities from India, China and USA.

\begin{figure}[!h]
	\centering
	\subfloat[Top 20 cities within each Countries]{
		\includegraphics[width=0.5\textwidth]{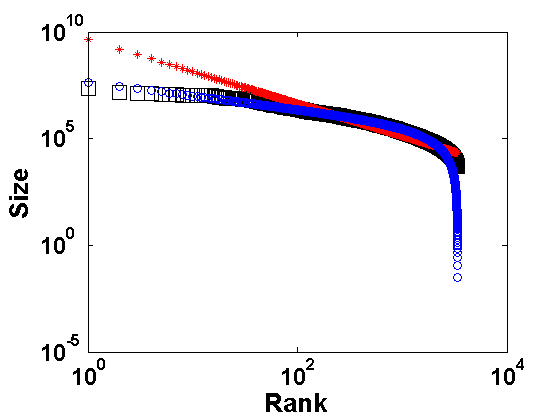}
		\label{FIG:WorldA}}
	\subfloat[Cities with size greater than 3 Lakhs]{
		\includegraphics[width=0.5\textwidth]{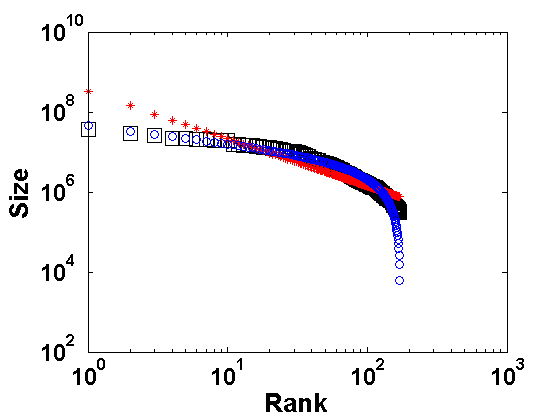}
		\label{FIG:WorldB}}
	\caption{Plot of the actual and the predicted sizes over ranks for
		 Worldwide city sizes (estimated) in 2018 (Black square: observed sizes, blue circle: RO fits, red star: Pareto fits)}
	\label{FIG:World}
\end{figure}

We combine the population of the above mentioned 3371 cities across the world 
and apply the proposed RO fitting along with the usual Pareto law. 
Since there are wide varieties of city sizes,  as expected, 
the Pareto modeling fails miserably (KS measure 7.474; $\nu=1.504$).
But, the universal RO approach, using the DGBD, again provides much better fit for the distribution of world cities
with an KS error measure of only 0.04 (Figure \ref{FIG:WorldA}); the parameters of the fitted DGBD are
$a=0.632$ and $b=1.963$. Note that, here we have a significantly larger values of $b$ 
indicating wider spectrum of the city sizes.

Due to the unavailability of any global  pattern for city size distribution, a commonly used practice is to study the distribution of city sizes, within the framework for Pareto law, considering the cities with sizes larger than a specific limit. 
Let us now  consider all cities (171) around the world having populations of at least 300000  in the year 2018 and  
 sizes of these 171 cities are  fitted by the DGBD and the Pareto distributions. 
As expected, Pareto modeling gives a reasonable fit to these high-end city sizes (KS=0.703, $\nu=1.178$)
but the proposed RO modeling, via DGBD, yields even a better fit than Pareto with a KS value of 0.041
($\widehat{a}=0.426$, $\widehat{b}=1.302$); 
see Figure \ref{FIG:WorldB} for the corresponding fitted sizes along with their original population.
This further supports our claim on the universality of the rank order modeling, using DGBD,  in analyzing any kind of city size distributions.

Finally, we end our illustrations with a remark on the modeling of the country sizes.
 We note that while analyzing the populations of 226 major countries ( population $>$ 5000) 
  across the world as per the 2018 estimate given in \cite{world2018},
the Pareto law again fails tremendously (KS=14.091, $\nu=2.257$) due to their wider spectrum.
On the other hand, the RO fit is again much better leading to the KS error of only $0.061$
($\widehat{a}=0.976$, $\widehat{b}=1.499$). This result justifies further the universal nature of the DGBD. 
Probably, the countries around the world have a similar distributional structure as that of the  cities within a country  
and we should be able to explain them as well with a city size law.

\section{Discrimination and Evolution of the City Size Distributions}
\label{SEC:discrimination}

\begin{figure}[!h]
	\centering
	\includegraphics[width=18cm]{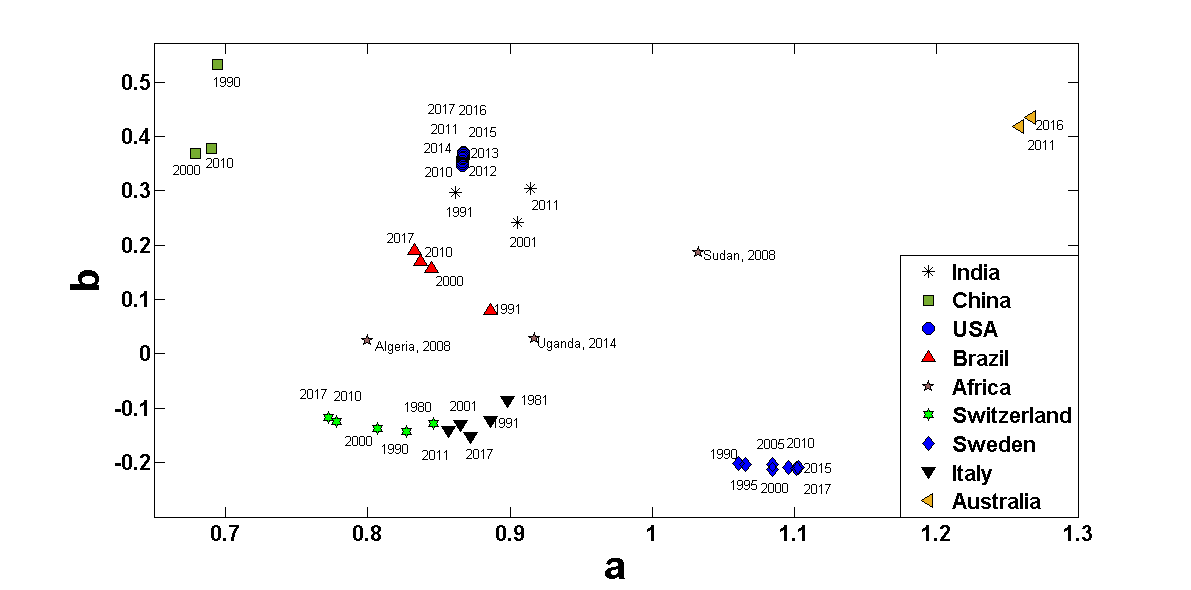}
	\includegraphics[width=18cm]{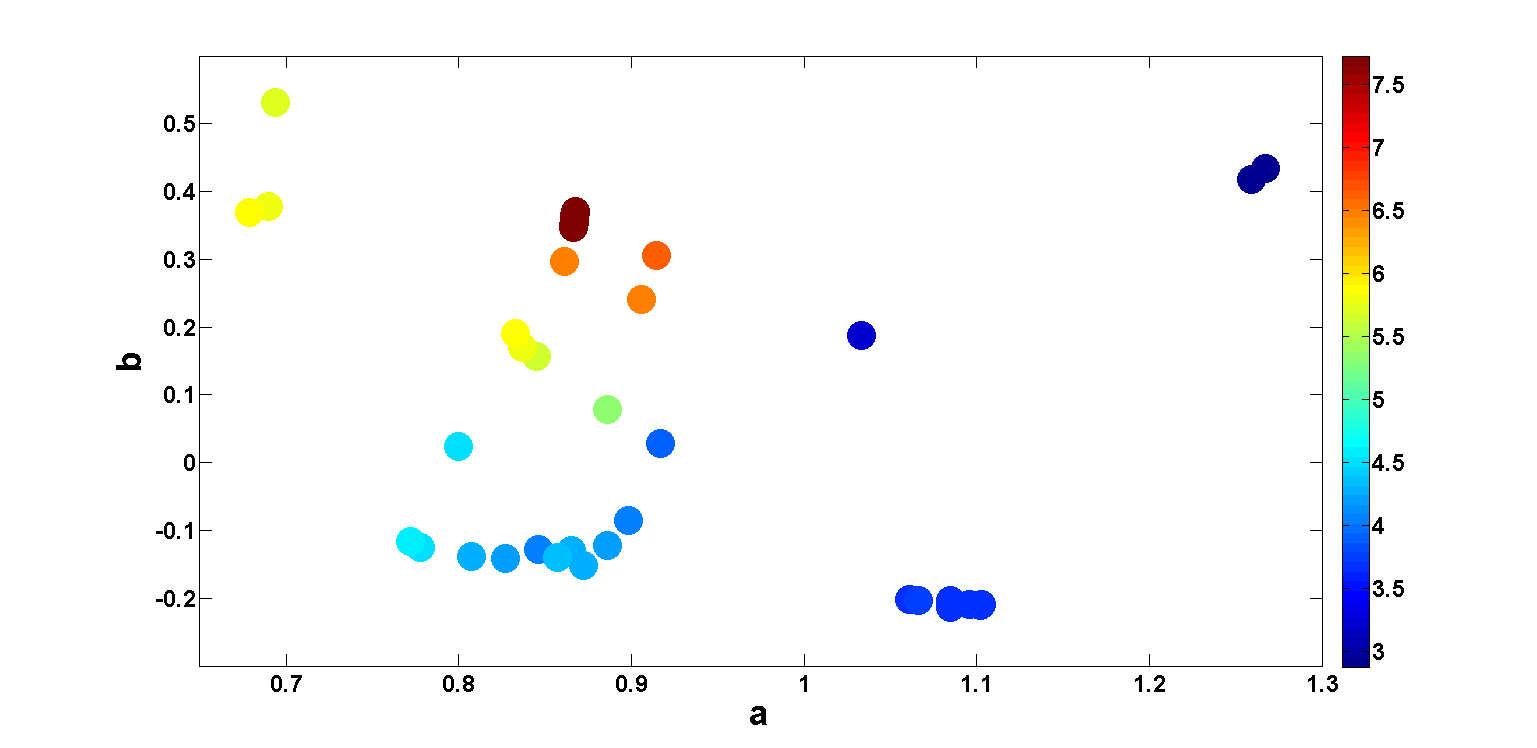}
	\caption{(Top) Estimated parameters $(a,b)$ of the fitted DGBDs for different countries and different years  and (Bottom) the corresponding entropy values. }
	\label{FIG:AB}
\end{figure}

An important usefulness of the particular RO distribution, namely the DGBD, 
is its ability to characterize the city size distribution 
through the two parameters $(a, b)$.
It is noted in Sec.~\ref{SEC:ROD} that these two parameters $(a,b)$  completely define the shape of 
the RO distribution (Fig.~\ref{FIG:RO_posA})
and hence the characteristics of the city sizes along with the entropy of the underlying process (Fig.~\ref{FIG:Entropy_N20}). 
Therefore, their estimated values in the cases of different countries 
enable us to characterize and discriminate the corresponding city-size distributions  
and the underlying uncertainty through its entropy.
Similarly, the changes observed in these estimated parameters at different time points for any fixed country 
also indicate the evolution of the corresponding process within that country 
in terms of the shape and the uncertainty of its city-size distribution.
For illustration, in Figure \ref{FIG:AB}(top), 
we have plotted the estimated values of $(a, b)$ for all our illustrative examples
with 11 countries across the seven continents at different years,
which provides a vividly clear comparison of their city-size distributions and their evolution.
The heat-map of the entropy values of the corresponding fitted DGBDs 
are also presented in Figure \ref{FIG:AB}(bottom) for entropic analyses. 

One can clearly see from Figure \ref{FIG:AB} that the large countries like India, China, USA or Brazil
have positive $b$ values indicating a  completely different class from the smaller but  developed countries like
Italy, Sweden or Switzerland with  slightly negative values of $b$.
Moreover,  the first group of countries clearly belong to a class of 
higher entropy and hence have more uniform city size distributions compared to that of the second group; 
but the spectrum of city sizes within the first group is much wider than those in the second group due to their larger number of cities.
Further, within the first group of countries, China has a clearly different city-size distribution
compared to those of India and US; the second group is structurally more similar.
China has the lowest entropy among these three with USA having the highest entropy value; 
this is because the urban population of USA is well distributed (most uniformly) within several cities. This pattern becomes less uniform for India and China,
where the city sizes are 
 more distinct. 
In terms of evolution through the years, both India and China have a drastic change in its city-size distribution 
between the twentieth and twenty-first centuries, which can be seen from differences in their $(a, b)$ estimates.  
In the twenty-first century, both China and USA become quite stable in terms of their city size distributions,
but that of India is still changing significantly in shape from the year 2001 to 2011 with a very little increase in entropy. 
The city-size distribution of Brazil is found to be in the same entropy range as China over the years with a small increasing trend,
but structurally coming closer to the India-USA cluster after an abrupt change in the last decades of twentieth century.

The city sizes of European countries like Italy, Sweden and Switzerland, on the contrary, 
have much smaller distribution spectrum  but greater variability in their city sizes.
They are completely separated from the group of above mentioned countries through its estimated values of $b$
which are (equal or)  very close to zero from the negative side due to the uniformity among the smaller 
cities within these countries (as already discussed in Section \ref{SEC:Europe}).
Among them, the city-size distribution of Italy and Switzerland are pretty similar 
although Switzerland has slightly greater entropy value due to greater uniformity in its cities' sizes. 
For both of these countries, the entropy is increasing with time indicating a possible growth mechanism of the cities of these  countries --
sizes of smaller cities are growing at a faster rate than those of the larger cities. 
Such a change is more prominent in Switzerland than in Italy where big cities like Milan and Rome is still dominating.
Sweden, on the other hand, has quite a different city size distribution with comparatively lower entropy as 
 handful number of large cities in Sweden accommodate  majority of its urban population.
 
African countries are slightly in between these two groups,
with Sudan having significantly lower entropy and randomness in its city size distribution followed by Uganda and Algeria, respectively. 
The urban populations of Algeria is comparatively well distributed between several cities (higher entropy),
whereas the majority of the urban population of Sudan is concentrated in few largest cities 
(45\% in top 3 cities, and 70\% in top 10 cities among a total of 63 cities).
The distribution of Australian Agglomeration, although structurally completely different from all other countries considered,
are similar to the city size distribution of  Sweden and Sudan;  a majority of its population is  concentrated only in a  few large UA 
(57\%  in top 6 UA among the total of 400 UAs). This leads to the lowest  possible entropy for their size distribution
characterized with the wide range of the spectrum of Australian human settlements having significantly distinct sizes.

This exhaustive study helps us to conclude that the DGBD, along with its entropy formulation, 
can characterize the evolution of the urbanization  and can also  discriminate 
between the city size distributions of different countries.  
 The  RO approach, with its two parameters, provides greater clarity for the characterization and  discrimination between the city size distributions of different countries than the one parameter power law. 



\section{Conclusions and Discussions}

There are many competing criteria in the study of cities. This has made the science of city planning such a challenging one. 
Despite the enormous complexity and diversity of human behavior and extraordinary geographic variability, 
we have shown that the size distribution of  cities around the  world follows a universal law. 
A two-parameter distribution,  with the parameter values varying within a very short range, 
can illustrate the empirical data for the city sizes across the world. 
Moreover, the entropy analysis presented here is used to quantify the variation in the distribution of city sizes. 
The underlying city size distribution of the countries with a wider spread of city sizes, i.e. with a few extremely populous cities along with several smaller cities,  can be characterized with low entropy. 
This is possibly connected with the lesser uncertainty in those country's human settlements 
focusing more on 
larger cities (e.g., Swedish cities, Australian UA).   
On the other hand, the uncertainty and internal movement within cities are 
expected to be higher for countries with almost equal city sizes
or with lower spread of city sizes; they further lead to comparatively higher  entropy of the underlying DGBD
(e.g., Switzerland, Italy or China).


Among several possible extensions of our work, 
the immediate one should be an in-depth analysis of the drivers of the city dynamics
within different countries. Here, we have already studied the discrimination and evolution of countries' city size distribution, through the entropy analysis and two fitted parameter values of the proposed RO distribution, the DGBD.
These two parameters can be thought of as two $sensors$  
of the city dynamics, those  could be further associated with observable economic, social or environmental factors;
these observables can then be controlled for any suitable planning of the city-sizes in a country as per the requirements.
Furthermore,  the possible linking of these two sensor parameters of city populations 
with the underlying entropy and the maximum entropy principle might shed further light on their physical significance.
Such an analysis may be explored with the inverse problem of characterizing the DGBD
as a maximum entropy distribution under appropriate constraints. 
We hope to pursue some of these interesting and useful extensions in future 
and believe that the proposed universal rank-order modeling for the city sizes 
(comprising the whole data set, starting from smaller human settlements to the most populous city in the country) 
will open up new avenues of research in this area.


\nopagebreak
\begin{acknowledgements}
The authors wish to thank the Editor and two anonymous referees for careful reading of the manuscript 
and several valuable constructive suggestions which have significantly improved the paper.
The work of the first author, AG, is supported by the INSPIRE Faculty Research Grant from 
the Department of Science and Technology, Government of India, India.
\end{acknowledgements}



\begin{thebibliography}{99}


\bibitem{pnas} L. M.A. Bettencourt, J. Lobo, D. Helbing, C. Kuhnert and G.B. West, Proc. Nat. Acad. Sci. {\bf 104} (2007) 7301.
\bibitem{batty} M. Batty et al., Science {\bf 319} (2008) 769 . 
\bibitem{zipf} G.K.Zipf, Human Behavior and the Principle of Least Effort (Addison-Wesley, Cambridge, MA, 1949). 
\bibitem{usa} D. Zanette and S.C. Manrubia, Phys. Rev. Lett. {\bf 79} (1997) 523 .
\bibitem{brazil} N. J. Moura Jr. and M. B. Ribeiro, Physica A {\bf 367} (2006) 441 .
\bibitem{kgbb} K. Gangopadhyay and B. Basu, Physica A {\bf 388} (2009) 2682 . 
\bibitem{kgbb2013} K. Gangopadhyay and B. Basu, in F. Abergel, B. Chakrabarti, A. Chakraborti, A. Ghosh (eds) 
{\it{Econophysics of Systemic Risk and Network Dynamics}}. New Economic Windows. Springer, Milano (2013). 
\bibitem{stanley2000} LAN Amaral, A. Scala, M. Barthélémy, H.E. Stanley   
 Proc. Nat. Acad. Sci. {\bf 97} (2000) 11149. 
 \bibitem{newman2005} M.J. Newman 
Contemporary Physics {\bf 46} (2005) 323. 
\bibitem{12kgbb} Y. Sasaki, H. Kuninaka, N. Kobayashi, and M. Matsushita, J. Phys. Soc. Jpn. {\bf 76(7)} (2007) 074801.
\bibitem{13kgbb} H. D. Rozenfeld, D. Rybski, J. S. Andrade Jr., M. Batty, H. E. Stanley, and H. A. Makse, 
Proc. Nat. Acad. Sci. {\bf 105} (2008) 18702.
\bibitem{14kgbb} H. A. Makse, S. Havlin, and H. Eugene Stanley, Nature {\bf 377} (1995) 608. 
\bibitem{15kgbb} H. Kuninaka and M. Matsushita, J. Phys. Soc. Jpn., {\bf 77} No.11 (2008) 114801.
\bibitem{16kgbb} L. C. Malacarne, R. S. Mendes, and E. K. Lenzi, Phys. Rev. E {\bf 65} (2002) 017106.
\bibitem{17kgbb}  B. B. Mandelbrot, The Fractal Geometry of Nature (Freeman, New York, 1977).
\bibitem{18kgbb}  J. Laherrere and D. Sornette, Eur. Phys. Jour. B {\bf 2} (1998) 525 .
\bibitem{19kgbb}  C. Tsallis, J. Stat. Phys. {\bf 52}, (1988), 479.
\bibitem{luck17I} J Luckstead et al., Physica A {\bf 474} (2017) 237.
\bibitem{luck17U} J Luckstead et al., Physica A {\bf 465} (2017) 573.
 \bibitem{Martinez-Mekler09}
G. Martínez-Mekler,  R.A. Martínez, M.B. del Río, R. Mansilla, P. Miramontes, and G. Cocho, 
PLoS One, {\bf 4(3)} (2009) e4791.
\bibitem{Ausloos/Cerqueti:2016}M. Ausloos, and R. Cerqueti, 
PloS one, {\bf 11(11)} (2016) e0166011.

\bibitem{Alvarez-Martinez/etc:2011}
R. Alvarez-Martinez, G. Martinez-Mekler, and G. Cocho, 
Physica A: Statistical Mechanics and its Applications, {\bf 390} (1) (2011) 120 


\bibitem{Alvarez-Martinez/etc:2014}
R. Alvarez-Martinez, G. Cocho, R.F. Rodríguez, and G. Martínez-Mekler,  
Physica A: Statistical Mechanics and its Applications, {\bf 402}, (2014), 198

\bibitem{Oscar2017} O. Fontanelli, P. Miramontes, G. Cocho and W. Li, Royal Soc. Open Science, {\bf 4} 170281 (2017)

\bibitem{Alvarez-Martinez/etc:2018}
R. Alvarez-Martinez, G. Cocho and G. Martinez-Mekler,  
 Chaos: An Interdisciplinary Journal of Nonlinear Science, {\bf 28}(7) (2018) 075515.

\bibitem{Lalit/etc:2018}
M. Lalit, A.Biswas, A. Ghosh, and D. Sengupta,  
ROSeq: A rank based approach to modelling gene expression in single cells
bioRxiv, doi:10.1101/374025 (2018).

\bibitem{jaynes} E. T. Jaynes,  
 Physical Review, {\bf 106}, (1957), 620 . 
\bibitem{kapurbook} J.N. Kapur,   Maximum-entropy models in science and engineering. John Wiley \& Sons (1989).
\bibitem{battygis} N. Mohajeri, Jon R. French and M. Batty, Annals of GIS, {\bf 19(1)}, (2013), 1.
\bibitem{desa} http://www.un.org/en/development/desa/news/population/
world-urbanization-prospects-2014.html 
\bibitem{un} https://www.un.org/development/desa/en/news/population/ 2018-revision-of-world-urbanization-prospects.html\\
Data is collected from the Population Division of the United Nations Department of Economic and Social Affairs (UN DESA).
\bibitem{census2011} http://www.censusindia.gov.in
\bibitem{world bank data} http://dataworldbank.org/data-catalog/world-development-indicators,2014
\bibitem{plos2007} E.H. Decker, A.J. Kerkhoff and M.E. Moses, PLoS ONE, {\bf 2(9)}, (2007) e394.  
\bibitem{india data} http://censusindia.gov.in
\bibitem{china data} http://www.citypopulation.de/China
\bibitem{usa data} https://www.census.gov/data/tables/2017/demo/ popest/total-cities-and-towns.html, \\https://www.citypopulation.de/USA.html
\bibitem{brazilcity} http://www.citypopulation.de/Brazil
\bibitem{africa} http://www.citypopulation.de/Sudan, http://www.citypopulation.de/Uganda, http://www.citypopulation.de/Algeria
\bibitem{italy} http://www.citypopulation.de/Italy
\bibitem{sweden} http://www.citypopulation.de/Sweden
\bibitem{switzerland} http://www.citypopulation.de/Switzerland
\bibitem{australia} https://www.citypopulation.de/Australia.html, https://www.citypopulation.de/Australia-Agglo.html,\\
https://www.citypopulation.de/Australia-AggloEst.html,\\ 
http://worldpopulationreview.com/countries/australia-population
\bibitem{world2018}http://worldpopulationreview.com/world-cities,
http://worldpopulationreview.com/us-cities
\bibitem{afr2017}African Economic Outlook 2017, African Development Bank.
\end{thebibliography}
\end{document}